\documentclass[showpacs,preprintnumbers,amsmath,amssymb]{revtex4}
\usepackage{epsfig}
\usepackage{psfrag}
\usepackage{amsfonts}
\usepackage{graphicx}
\usepackage{dcolumn}
\usepackage{bm}
\usepackage{xcolor}
\usepackage{mathtools}

\begin{document}

\title{Axion-Polaritons in quark stars: a possible solution to the missing pulsar problem}

\author{E. J. Ferrer and V. de la Incera}
\affiliation{Department of Physics and Astronomy, University of Texas Rio Grande Valley, 1201 West University Dr., Edinburg, TX 78539}

\date{\today}

\begin{abstract}
This paper  proposes an alternative mechanism to solve the so-called missing pulsar problem, a standing paradox between the theoretical expectations about the number of pulsars that should exist in the galaxy center of the Milky Way and their absence in the observations. The mechanism is based on the transformation of incident $\gamma$ rays into hybridized modes, known as axion-polaritons, which can exist inside highly magnetized quark stars with a quark matter phase known as the magnetic dual chiral density wave phase. This phase, which is favored over several other dense matter phases candidates at densities a few times nuclear saturation density, has already passed several important astrophysical tests. In the proposed mechanism, the absence of young magnetars occurs because as electromagnetic waves inside the star can only propagate through the hybridized modes, incident photons coming from a $\gamma$-ray burst get transformed into massless and massive axion polaritons by the Primakoff effect. Once thermalized, the massive axion-polaritons can self-gravitate up to a situation where their total mass overpasses the Chandrasekhar limit for these bosons, producing a mini blackhole that collapses the star.

\end{abstract}

\maketitle
\section{Introduction}

An astronomical puzzle called the missing pulsar problem (MPP) refers to the failed expectation to observe a large number of pulsars within the distance of 10 pc of the Milky Way galaxy center (GC). Theoretical predictions have indicated that there should be more than $10^3$ active radio pulsars in that region \cite{G-M-P}, but these numbers have not been observed. This paradox has been magnified by the observation of magnetar SGR J1745-29 \cite{Magnetar-Detect} by the NuSTAR and Swift satellites. This observation revealed that the failures to detect ordinary pulsars at low frequencies could not be simply due to strong interstellar scattering but to an intrinsic deficit produced by other causes. Furthermore, as pointed out in \cite{Dexter}, the detection of the young ($T \sim 10^4$ yr) magnetar SGR J1745-29 implies high efficiency for forming magnetars from massive stars in the galaxy center because it will be barely likely to see a magnetar unless magnetar formation is efficient there. It was then argued that the efficiency in magnetar formation could be elicited by an unusual progenitor population in the GC and that the MPP can be explained as a consequence of a tendency to create short-lived magnetars rather than long-lived ordinary pulsars. 

On the other hand, the Milky Way GC is a very active astrophysical environment with numerous $\gamma$-ray emitting point sources \cite{GR}. The sources of $\gamma$-ray-burst (GRB) show an isotropic distribution over the whole sky flashing with a rate of 1000/year. The energy output of these events is $\sim 10^{56} - 10^{59}$ MeV, with photon energies of order $ 0.1 - 1$ MeV \cite{GRB}.  Each one of these events can then produce at least $10^{56} - 10^{59}$ $\gamma$-photons. If we assume that about $10 \%$ of these photons can reach a magnetar in the GC, which is a conservative estimate if the star is in the narrow cone of a GRB beam, then at least about $10^{55} - 10^{58}$ of those photons could reach the star per each GRB event. 

In this paper we propose a possible mechanism to explain the MPP. The mechanism is an application to astrophysics of the fundamental theoretical results found in \cite{EJF-ViC-NPB} about matter-light interactions in a magnetized dense quark matter phase known as the magnetic dual chiral density wave (MDCDW) phase. The mechanism is based on the following assumptions: (1) The GC is mostly populated by short-lived magnetars, whose internal structure is that of a quark star with inner magnetic fields of order $10^{17}$ G or higher. (2) The quarks inside the star are in the MDCDW phase (3) The GC region under consideration is very active with GRB. We shall show that under these conditions, the original magnetars can readily collapse into mini blackholes, thereby producing a pulsar desert.

To understand how this is possible, one should first call attention to a large number of studies (For a review see Ref. \cite{ICReview}) that have found that spatial inhomogeneous phases are favored in Quantum Chromodynamics (QCD) and in effective theories of QCD in the region of intermediate densities and relatively low temperatures, where the interior of the neutron stars (NSs) lie. In particular, many works have found that at intermediate densities, matter phases that feature spatially inhomogeneous quark-hole condensates are energetically favored over the chirally symmetric phase or over homogenous phases with broken chiral symmetry \cite{largeNQCD}-\cite{ferrer-incera-sanchez}. The MDCDW phase is an example of a chirally broken inhomogeneous phase with a quark-hole condensate in the presence of a magnetic field \cite{KlimenkoPRD82}. What makes this phase particularly interesting is the non-trivial topology of the fermion dynamics that gives rise to anomalous electric transport properties \cite{Topological-Transport-1, Topological-Transport-2}, stability against fluctuations at finite temperature \cite{Ferrer-Incera19}, and matter-light interaction effects with the consequent creation of axion-polaritons (APs) \cite{EJF-ViC-NPB}.

The MDCDW phase has been shown to be energetically preferred over the homogeneous chiral phases (broken and unbroken) in a wide range of the parameter space \cite{MDCDW-3}. At sufficiently strong magnetic fields ($\gtrsim 2\times 10^{18}$G), the condensate is favored at all the intermediate densities and up to temperatures of tens of MeV, conditions relevant for old NSs. In addition, this phase has been found to be consistent with a series of astrophysical constraints. For example, the maximum stellar mass of a hybrid star with a quark matter core in the MDCDW phase \cite{InhStars} was shown to be compatible with known observed stellar mass maxima ($M \gtrsim  2M_{\odot}$) \cite{Demorest, Antoniadis}. Furthermore, at strong magnetic fields, the speed of sound in this phase has been shown to reach values beyond the conformal limit \cite{Aric-2}, in agreement with NS expectations \cite{vs}. Finally, the heat capacity of an NS core made of MDCDW matter has been shown \cite{PRD20} to be well above the lower limit expected for NSs ($C_V\gtrsim 10^{36}(T/10^8)$ erg/K) \cite{Cv-NS}. These results indicate so far that the MDCDW phase is a viable candidate for the inner phase of NSs with strong internal magnetic fields. 

Let us outline, in a nutshell, how the MSP is solved when the assumptions outlined above are realized. Assumptions (1) and (2) means that the pulsars in the GC are mostly quark stars with large inner magnetic fields, and with their quark matter in the MDCDW phase. Assumption (3) states that these quark-magnetars are bombarded by $\gamma$ rays, a natural condition in the GC. Under these conditions, incident $\gamma$ rays will be converted into APs through the Primakoff effect \cite{Primakoff}. The APs are the propagation modes of electromagnetic waves inside the MDCDW medium. They are hybridized modes of photons and axions because in the MDCDW theory, matter-light interactions are described by the equations of Axion Electrodynamics \cite{EJF-ViC-NPB}. There are two APs modes, one massless and one massive.  The massive AP has a mass large enough to be trapped inside the star by self-gravitation, as we will see in more detail in subsequent sections. If the number of generated APs that self gravitate in the star's center is higher than the Chandrasekhar limit for these bosons, the AP's create a mini black hole in the center of the star that destroys the host NS. We will show that for  the conditions under consideration and for the parameter values characteristic of these compact objects, the star collapse under this mechanism is unavoidable.

The paper is organized as follows: In Sec. II, to facilitate reader’s understanding, we review the main equations that lead to the existence of the AP modes in the MDCDW phase \cite{EJF-ViC-NPB}. In Sec. III, we analyze the possible conversion of photons into gapped APs in the interior of a NS in the MDCDW phase for the cases of hybrid and quark stars. Then, in Sec. IV, we discuss how the APs once created by a GRB inside a NS can accumulate in the star center until they exceed the Chandrasekhar limit thus producing a mini blackhole. Sec. V contains our concluding remarks.
 
\section{The axion-polariton modes in magnetized dense quark matter}\label{S2}

Neutron Stars have cores in a region of intermediate-to-large densities (i.e. about a few times the nuclear saturation density), where quarks could be deconfined.  On the other hand, magnetars are NSs with surface magnetic fields $\sim10^{15}$ G, and
with inner fields even stronger reaching values of order a few times $10^{17}$ G \cite{relativisticmagnetohydro}.  Magnetars' densities and inner magnetic fields are compatible with the realization of the MDCDW phase. This phase is characterized by a spatially inhomogeneous chiral condensate $M(z)= \Delta e^{iqz}$, which spontaneously breaks chiral and translational symmetries. The condensate's magnitude $\Delta=-m/2G$ and modulation $q$ are dynamical parameters that have to be found by minimizing the effective action of the theory \cite{KlimenkoPRD82}. 

An important feature of the MDCDW phase is the presence of the chiral anomaly $(\kappa/8)\theta F^\ast_{\mu\nu} F^{\mu\nu}$, with $\kappa=\frac{2\alpha}{\pi m}$, in the free energy \cite{Topological-Transport-1,Topological-Transport-2}. The chiral anomaly couples the electromagnetic strength tensor and its dual to the axion field $\theta(\mathbf{r})=qmu(\mathbf{r})$ that is proportional to the phonon fluctuation $u(\mathbf{r})$ of the condensate phase \cite{EJF-ViC-NPB}. Such a term is responsible for various topological effects: anomalous electric charge; anomalous Hall current \cite{Topological-Transport-2}; and mixing between the axion field and the photon that makes the propagation of electromagnetic waves into this medium highly non-trivial  \cite{EJF-ViC-NPB} as described below.  

Let us consider phonon fluctuations in a system of strongly interacting quarks at finite density in the MDCDW phase with a static and uniform background magnetic field $\mathbf{B}_0$. If this matter is bombarded by linearly polarized electromagnetic waves with their electric field parallel to $\mathbf{B}_0$, the coupling between the phonons and the photons in the chiral-anomaly term is linear and the linearized field equations in the neutral medium take the form \cite{EJF-ViC-NPB},
\begin{eqnarray} \label{Wave-Eq}
&\frac{\partial^2\mathbf{E}}{\partial t^2}=\mathbf{\nabla}^2 \mathbf{E}+\frac{\kappa}{2} \frac{\partial^2\theta}{\partial t^2} \mathbf{B}_0
\\
&\frac{\partial^2\theta}{\partial t^2}- v_z^2 \frac{\partial^2\theta}{\partial z^2}-v_\bot ^2 (\frac{\partial^2\theta}{\partial x^2}+\frac{\partial^2\theta}{\partial y^2})+\frac{\kappa}{2} \mathbf{B}_0 \cdot \mathbf{E}= 0.
\end{eqnarray} 

which in momentum space are given as
\begin{equation}\label{P-Space-1}
\left (\omega^2-p^2 \right ) E-\left (\frac{\kappa}{2} \omega^2 B_0 \right ) \theta=0
\end{equation}

\begin{equation}\label{P-Space-2}
-\left (\frac{\kappa}{2}B_0 \right ) E+\left (\omega^2-P^2 \right )\theta=0
\end{equation}
where 
\begin{equation}\label{P}
P^2=v_z^2p_z^2+v_\bot^2p_\bot^2.
\end{equation}
with $v_z$ and $v_\bot$ being the parallel and transverse group velocities. Their squares are given by linear combinations of the Ginzburg-Landau free energy expansion coefficients times powers of the dynamical parameters \cite{Ferrer-Incera19}. In the sixth-order expansion, they take the form 
\begin{equation}\label{Parallel-Coef}
v^2_z= a_{4,2}+m^2 a_{6,2} +6q^2a_{6.4}+3qb_{5,3},
\end{equation} 
\begin{equation}\label{Transverse-Coef}
v_\bot^2= a_{4.2}+m^2 a_{6.2}+2q^2 a_{6.4}+qb_{5,3},
\end{equation} 
The $a$ and $b$ coefficients are functions of the magnetic field, baryonic chemical potential, and temperature. Explicit expressions for them can be found in Ref. \cite{MDCDW-3}. 

Solving the dispersion relation obtained from (\ref{P-Space-1})-(\ref{P-Space-2}), we have
\begin{equation}
\det
  \left[ {\begin{array}{cc}
   \omega^2-p^2  &-\kappa\omega^2 B_0/2  \\
   -\kappa B_0/2 &\omega^2-P^2  \\
  \end{array} } \right]=0
\end{equation}
From where,we obtain two hybridized propagating modes known as axion polaritons \cite{EJF-ViC-NPB}, one gapless and one gapped
\begin{equation}\label{Frequencies}
\omega^2_{0}=\omega_1^2-\omega_2^2, \quad \omega^2_{\delta}=\omega_1^2+\omega_2^2
\end{equation}
with
\begin{equation}\label{A}
\omega_1^2=\frac{1}{2}[p^2+P^2+(\frac{\kappa}{2} B_0)^2],
\end{equation}
\begin{equation}\label{B}
\omega_2^2=\frac{1}{2}\sqrt{ [p^2+P^2+(\frac{\kappa}{2} B_0)^2]^2-4p^2P^2}.
\end{equation}

This means that electromagnetic waves can only propagate inside the MDCDW medium as APs.

Notice that the AP gap
\begin{equation}\label{AP-Mass}
\omega_{\delta}(\vec{p}\rightarrow0)=\delta=\alpha B_0/\pi m
\end{equation}
 is proportional to the magnitude of the applied magnetic field. For a magnetic field value of  $B\simeq10^{17}$ G, $\delta$ is in the range $[0.06, 0.5]$ MeV and $m \in [23.5, 2.8]$ MeV for intermediate baryonic densities $\rho \sim 3 n_s$  \cite{MDCDW-3}.

\section{Conversion of photons into axion polaritons inside magnetars}\label{S3}

Based on the results outlined in Sec. \ref{S2}, if the interior of magnetars hosts quarks in the MDCDW phase, $\gamma$-photons that reach the quark matter core will propagate inside the medium as axion polaritons. The conversion of a large number of $\gamma$-photons into axion polaritons is possible through the Primakoff effect \cite{Primakoff}. In the context of the MDCDW dense quark matter, the Primakoff effect allows incident photons to get transformed into axion polaritons thanks to the anomalous axion-two-photons vertex and the existence of a background magnetic field. This effect can produce many gapped APs. As the APs are bosons, once thermal equilibrium is reached, they will be gravitationally attracted to the center of the star, where they will accumulate in large numbers.

In what follows, we will analyze whether and under which conditions a NS under an intense $\gamma$-ray bombardment could acquire a number of AP's large enough to affect the star's stability. 

\subsection{$\gamma$-ray attenuation in neutron stars}

  The first question we need to elucidate is whether the $\gamma$-ray radiation produced in a GRB will be able to reach the quark-matter region of the NS. The $\gamma$-rays are attenuated in a medium mainly through their interaction with electrons. As it is known, three processes are driving the attenuation: the photoelectric effect, pair creation, and Compton scattering. The photoelectric effect works for electrons that are tight to atoms, but in NS, the electrons are already stripped out from the atoms leaving at most ions and free electrons. For pair creation, the $\gamma$ photons would need energies higher than $2m_e=1.2$ MeV, which is beyond that of the most energetic photons produced in the majority of GRB. Therefore, the overall attenuation process is mainly through the Compton effect. 

The radiation intensity attenuation can be calculated from the formula
\begin{equation}\label{Attenuation-Formula}
 I=I_0 e^{-\sigma n_e L}
  \end{equation}
where $I_0$ is the incident radiation intensity, $I$ the intensity at a thickness L inside the medium, $\sigma$ the cross-section and $n_e$ the electron number density, since the electrons, being the star component with the lower mass, are the particles with a higher contribution to Compton scattering. The cross-section $\sigma$ is calculated for the Compton effect between photons of certain energy and the free electrons.

Quark matter can occur in principle in two classes of NS: hybrid stars and quark stars. In hybrid stars, as the density increases from the surface to the inner region, different states of matter are realized \cite{NS Structure}. Iron nuclei and free electrons mainly populate the outer crust, while in the inner crust, once the density overtakes the neutron drip density $n_d=4 \times 10^{11}$ $g/cm^3$ where the neutron chemical potential is zero, neutrons leak out of nuclei, and at its highest densities most of the matter resides in a neutron fluid rather than in nuclei. Then, in the several kilometers thick of the outer core, at densities in the range  $0.5 n_s \leqslant n \leqslant 2 n_s$, the atomic nuclei dissolved into their constituents $n$ and $p$ (with a fraction rate 8:1), with a number of electrons roughly the same as the number of protons to secure neutrality, plus some other baryon resonances and mesons. Finally, in the inner core, at densities several times $n_s$, baryons are so close that they are smashed together, leading to quark deconfinement. Therefore, to reach the quarks in a hybrid star, the $\gamma$ photons will have to travel several kilometers through all these layers of different matter structures. 

Quark stars, on the other hand, have quite a different structure. A quark star can be made exclusively of strange matter, so it is also called strange star. The idea of a strange star was proposed by the Bodmer–Terazawa–Witten hypothesis \cite{BTW} based on the fact that strange matter has a lower energy per baryon than ordinary nuclei, including $^{56}Fe$. Thus, the actual ground state of the hadrons may be a strange matter. Later on, the equilibrium composition and the equation of state for strange matter were studied by other authors \cite{Farhi, Alcock, Paczynski, Weber}. The conclusion was that a strange star is formed by an absolutely stable phase consisting of roughly equal numbers of up, down, and strange quarks plus a smaller number of electrons (to guarantee charge neutrality). However, more recently \cite{Holdom}, it has been found, using a phenomenological quark-meson model that includes the flavor-dependent feedback of the quark gas on the QCD vacuum, that $u$-$d$ quark matter is, in general, more stable than strange quark matter. It can even be more stable than ordinary nuclear matter at a sufficiently large baryon number.

Quark stars exhibit a macroscopic quark-matter surface that is shrouded with an electron cloud that extends several hundred fermis above it \cite{Alcock}. As electrons are tied by the weaker electromagnetic force, they can go beyond the quark surface, which is held by the stronger nuclear force. Accordingly, the quark-star surface acts as a membrane that allows only ultrarelativistic matter to escape: photons, neutrinos, electron-positron pairs, and magnetic fields. No baryons can penetrate this membrane without being converted to quarks \cite{Paczynski}. Hence, for incoming $\gamma$ rays to reach the quark matter region in a quark star and put in action the Primakoff effect, they have to penetrate the electron cloud without a considerable attenuation loss.

Below, we analyze the attenuation of incoming $\gamma$ rays for the two NS cases, assuming they have quark matter in their interiors.

\begin{center}
\textbf{Hybrid stars}
\end{center}


As mentioned above, to reach the quark matter in the inner core of a hybrid star, the $\gamma$ radiation has to cross the crust and the outer core. Let us find the decay rate inside the outer core for different lengths. From Eq. (\ref{Attenuation-Formula}), in order to do this we need to know the cross-section $\sigma$ and the electron number density $n_e$ in that region. 

We first calculate the cross section associated to the Compton scattering of the $\gamma$ photons with the electrons in the outer core. The formula that gives the Compton scattering cross section is known as the Klein-Nishina formula \cite{Klein-Nishina, Longair} and is given by
 \begin{eqnarray}
\sigma&=&\frac{3e^4}{48\pi \epsilon_0^2 m_e^2 c^4}  \left [\frac{1}{x}\left (1-\frac{2(x+1)}{x^2} \right ) ln (2x+1)+\frac{1}{2x}+\frac{4}{x^2}-\frac{1}{2x(2x+2)^2} \right ]
\nonumber
\\
&=&2.49 \times 10^{-25}\left [\frac{1}{x}\left (1-\frac{2(x+1)}{x^2} \right ) ln (2x+1)+\frac{1}{2x}+\frac{4}{x^2}-\frac{1}{2x(2x+2)^2} \right ] cm^2,
\label{K-N Formula}
\end{eqnarray}   
where $x=\hslash \omega / m_e c^2$, is the ratio between the photon energy and the rest energy of the electron, that in our case taking for the maximum photon energy $\hslash \omega \approx 1$ MeV, we have $x\approx2$ and then the cross section reduces to
\begin{equation}\label{cross-section-value}
\sigma\approx 2.58 \times 10^{-25} cm^2
  \end{equation}

To calculate the electron number density, we consider that the ratio of neutrons to protons in this region is 8:1. Hence, taking into account that from the neutrality condition, the number of electrons should be equal to the number of protons we find 
\begin{equation}\label{N-e}
n_e=2n_s/8m_n,
  \end{equation}
where we considered that the density in the outer core is approximately twice the saturation density and $m_n$ is the neutron mass. Thus, in this case, $n_e=0.42\times 10^{38} cm^{-3}$. 

Hence, using the radiation intensity attenuation formula (\ref{Attenuation-Formula}), we have
\begin{equation}\label{N-e}
L=0.93 \times ln\left (\frac{I_0}{I} \right ) fm
  \end{equation}

In Table \ref{Table-1}, we give some values of the penetration depth corresponding to ratios of the incident to attenuated $\gamma$-ray intensities. We can observe that in a few fm, the intensity decays tremendously, while to reach the star core the $\gamma$ ray would have to penetrate several kilometers.

\begin{table}[ht] 
 \caption{$\gamma-ray$ intensity decay versus penetration distance in hybrid stars.}
\begin{center}
  \begin{tabular}{ | c  | c  |  c  |  c  |  c  | } 
  \hline
$I/I_0$ & $10^{-30}$ & $10^{-40}$ & $10^{-50}$ & $10^{-60}$\\ 
\hline
L(fm) & 64.2 & 85.7 & 107 & 128 \\ 
   \hline
  \end{tabular}
  \end{center}
  \label{Table-1}
 \end{table}

We conclude that in this case the $\gamma$ rays cannot reach the quark core to produce the AP's. Thus, the high electron density in the outer and inner crust prevent the creation of AP's in hybrid stars under $\gamma$-ray bombardment. 

\begin{center}
\textbf{Quark stars}
\end{center}


The situation with quark stars is different. Now the $\gamma$ rays have to cross only an electron cloud a few hundred fermi thick to reach the quark matter  \cite{Alcock}. As discussed above, the damping in the $\gamma$-ray intensity is produced by the Compton effect, with a cross-section value given by Eq. (\ref{cross-section-value}), since the energy of the $\gamma$ rays is the same as the one considered in the last case. What makes the difference now is the electron number density in the external star cloud.

The formula to calculate the electron number density beyond the quark surface was obtained in \cite{Alcock} and is given by
\begin{equation}\label{N-e}
n_e=\frac{9.49 \times 10^{35}cm^{-3}}{\left [ 1.2 \left (z/10^{-11} cm \right ) +4 \right ]^3} 
  \end{equation}
with $z$ being the height above the quark surface. The density increases as z decreases, meaning as it approaches the quark surface.

To be conservative, we should estimate the maximum attenuation. With that in mind, we assume that the electron density is constant across the electron cloud and equal to its maximum value, which occurs at z=0, just above the quark surface. We also know that the depth of the cloud should be just a few hundred fermis, but again to be on the conservative side, we will consider depths of up to 900 fm and estimate the attenuation rate right at the quark surface. Using these assumptions and  Eqs. (\ref{Attenuation-Formula}), (\ref{cross-section-value}) and (\ref{N-e}), we obtain the attenuation rates shown in Table \ref{Table-2}.

\begin{table}[ht]
 \caption{$\gamma-ray$ intensity decay versus penetration depth in quark stars.}
\begin{center}
  \begin{tabular}{ | c  | c  |  c  |  c  |  c  | } 
  \hline
$I/I_0$ &0.89 & 0.83 & 0.77 & 0.71\\ 
\hline
L (fm) &300 & 500 & 700 & 900 \\ 
   \hline
  \end{tabular}
  \end{center}
  \label{Table-2}
\end{table}

These results show that the attenuation of the $\gamma$ rays is minimal, an indication that if the GRB bombards the quark star, the Primakoff effect can become operative and convert $\gamma$-ray photons into AP in the star interior. The gapped AP produced via this mechanism will be attracted by gravity once in thermal equilibrium to be stored in the star center. 

There are other scenarios in the literature where instead of the bare-quark-star model here considered, a thin crust of different matter components surrounding the strange matter is assumed. The possibility of a thin crust of nuclear matter was suggested \cite{Alcock, NM-Crust} soon after the idea of strange starts emerged. Later, however, it became clear \cite{Usov} that since the temperature at the moment of the star formation is very high (a
few times $10^{11}$ K \cite{Inner-T} in the stellar interior), the rate of neutrino-induced mass
ejection is extremely high \cite{Neutrino} and, consequently, a few seconds after the star's formation the crust is completely blown away, leaving behind a bare strange-star surface. It was subsequently shown that the strange star remains nearly bare as long as its surface temperature is
higher than $\sim 3 \times 10^7$ K \cite{Usov-2} , a realizable value after the cooling by photon emission from the surface \cite{Photon-cooling}. 

On the other hand, older strange stars can be cold enough ($T < 3 \times 10^7$ K) to form a crust of
nuclear matter via mass accretion from the interstellar medium or from the leftover of the supernova explosion that originated the strange star. For this crust to be stable, a strong electric field created between the electron cloud and the net positive charge of the $u$ and $d$ quarks, which are closer to the quark surface (due to the migration of massive $s$ quarks to the star interior) should compensate the gravitational pull. Moreover, the matter density in the base of this crust should be smaller than the neutron drip velocity
 ($n_n^{drip} \simeq 4.3 \times 10^{11}$  g/cm$^3$). Otherwise, the neutrons could freely fall into the stellar core, and there they will be dissolved into their constituent quarks due to the absolute stability of the strange quark matter. This condition sets an upper limit for the crust thickness of the order of a few hundreds of meters \cite{NM-Crust}.
In this context, a new alternative for the crust composition of old and cold quark stars has been recently proposed. For this new description to work, the surface tension of the interface between the quark matter and the vacuum needs to be less than a certain critical value \cite{strangelets, Surface-Tension}, so that strange matter can be fissioned into smaller nuggets (strangelets)  of positively charged strange matter that will form a heterogeneous phase composed of a Coulomb lattice of positively charged strangelets immersed in a neutralizing sea of electrons \cite{strangelets}. Taking into account the effect of the surface tension and the Debye screening of the electric charges, the strangelet crust's size was found to range between zero and hundreds of meters thick  \cite{Alford}. If this alternative is realized, incoming $\gamma$-rays will meet quark matter already in the crust.

As it will become apparent in the following section, we are interested in young quark stars; hence our discussion will remain focused on bare quark stars.

\section{Accumulation of axion polaritons in the star interior and mini blackhole formation}\label{S4}

After confirming the viability to convert incoming $\gamma$-rays into AP's inside the quark star through the Primakoff effect, we need to consider how they can be accumulated inside the star center in numbers sufficiently large to produce a mini blackhole able to destroy the whole star. With that aim, we now turn our attention to the Chandrasekhar limit that determines the number of self-gravitating AP required to form the blackhole and induce the star collapse. For boson particles, this limit is given by \cite{Ch-Bosons-1, Ch-Bosons-2, Dermott}
\begin{equation}\label{N-Cha}
N_{AP}^{Ch}= \left(\frac{M_{pl}}{\delta}\right)^2=1.5 \times 10^{44} \left(\frac{MeV}{c^2\delta}\right)^2
\end{equation}
where $M_{pl}=1.22\times 10^{19}$ GeV is the Planck scale. For the range of $\delta$'s considered at the end of Section 2, we have that $N_{AP}^{Ch} \sim 10^{45}-10^{46}$.
 
Next, we need to check if the APs created by $\gamma$-rays in the energy range of $0.1-1$ MeV can escape the star gravitational attraction. To this aim we compare the escape velocity $v_e/c$ of the star
\begin{equation}\label{Axion-Max-Vel}
v_e/c=\sqrt{2GM_{star}/c^2R_{star}}, 
\end{equation}
which for a star of mass $M_{star}=2 M_\odot$ and radius $R_{star}=10$ km is $v_e/c\sim 0.8$, with the velocity $v_{AP}$ that the gapped APs can gain from incident gamma-rays. This velocity is given by
\begin{equation}\label{Axion-Max-Vel}
v_{AP}/c=\sqrt{1-\left( \frac{\delta \cdot c^2} {E_\gamma} \right)^2}. 
\end{equation}
where $E_\gamma$ is the energy the AP can gain from the $\gamma$-photons. We assume that the maximum value for $E_\gamma$ is $0.5$ MeV, which comes from considering an equal distribution of the $\gamma$-photon energy of $1$ MeV between the massless and massive APs. From (\ref{Axion-Max-Vel}) we see that for $E_\gamma \leqslant 0.5$ MeV, there is always an interval of AP-masses lying in the range $[0.06, 0.5]$ MeV for which the condition $v_{AP} < v_e$ is satisfied and hence all these APs will be trapped. For instance, for $\delta =0.4$ MeV$/c^2$ and $E_\gamma=0.5$ MeV we have $v_{AP}/c=0.6 < v_e/c=0.8$.
The $2 M_\odot$ here used is motivated by the findings of Ref. \cite{Nature}, which combined astrophysical observations and theoretical ab-initio calculations in a model-independent way, to conclude that deconfined quark-matter is the most plausibly state to form the cores of the heaviest reliably observed NSs with masses in the interval $2 M_\odot \leqslant M_{star} \leqslant 2.25 M_\odot$; while for $M_{star} \sim 1.4 M_\odot$ a hadron phase is the compatible one.
 Hence, a big portion of the gapped polaritons generated by the GRB will not escape the star. 
 
 Now, due to collisions with the quarks in the medium the AP's will loose kinetic energy, eventually attaining thermal equilibrium. The temperature at which that happens can be estimated by equating the thermal energy to the APs' kinetic energy. For instance, for APs with total relativistic energy $E_\gamma= 0.5$ MeV and rest mass $\delta= 0.4$ MeV/$c^2$, that temperature is of order  $\sim 10^8$K. At this point, they are drifted toward the star's center by the gravitational pull and start accumulating within a region of radius $r_{th}$, known as the thermal radius \cite{Ch-Bosons-2}, which is defined as the radius of the sphere where the thermal and gravitational energies of the particles in the stellar medium are equal,
\begin{equation}\label{Thermal radius-1}
(3/2) k_B T = G \frac{\delta \rho V}{r},
\end{equation}
which for a star matter density of about three times the nuclear saturation density can be written as
\begin{equation}\label{Thermal radius-2} 
r_{th} \sim \left (\frac{9k_BT}{8\pi G \rho \delta}\right )^{1/2}=241m \left (\frac {T}{10^7 K} \right )^{1/2} \left (\frac{10 MeV}{c^2 \delta} \right )^{1/2}
\end{equation} 
In the r.h.s of (\ref{Thermal radius-2}), we considered a reference temperature $T=10^7$K, which is the threshold for $\gamma$-radiation penetration as discussed above. Using the AP rest energy of $0.4$ MeV as before, we find $r_{th}= 3.8$ km, for $T=10^8$K; $r_{th}=1.2$ km, for $T=10^7$K; and $r_{th}= 0.1$ km, for $T=10^5$K. This range of temperatures are characteristics of young stars' temperatures, as for instance the Crab Nebula \cite{NST}, which is 1000 years old or less and have $T \lesssim 10^6$K. As shown above, for the scales of $\gamma$ photon's energies considered, most of the APs are trapped and besides, they cannot decay  in the MDCDW medium because as it was shown in Ref. \cite{EJF-ViC-NPB} its number density is conserved. As a consequence, the number of APs generated at $T\geqslant10^7$K will remain unchanged inside the star as T decreases. 

The question now is to estimate how many APs should concentrate in a sphere of radius $r_{th}$ to be self-gravitating and whether that number is equal or larger than the Chandrasekhar limit $N_{AP}^{Ch}$ to produce the collapse. To be self-gravitating the total mass of the APs within that sphere has to exceed the stellar baryonic mass of that volume \cite{Dermott}
\begin{equation}\label{Self-Gravitation} 
(N^{sg}_{AP})\delta \geqslant \rho V=\frac{4}{3}\pi r_{th}^3 \rho.
\end{equation} 
Solving for $N^{sg}_{AP}$ and using (\ref{Thermal radius-2}), the number of self-gravitating APs satisfies
\begin{eqnarray}
N^{sg}_{AP}& \geqslant&\frac{4 \pi}{3} (241 m)^3 \left (\frac{c^2 \rho}{10 MeV} \right )  \left (\frac{T}{10^7 K}\right ) ^{3/2} \left (\frac{10 MeV}{c^2 \delta} \right )^{5/2}
\nonumber
\\
&=&2.2 \times 10^{54}  \left (\frac{T}{10^7 K}\right ) ^{3/2} \left (\frac{10 MeV}{c^2 \delta} \right )^{5/2}=6.9\times 10^{57}\left (\frac{T}{10^7 K} \right )^{3/2}
\label{K-N Formula}
\end{eqnarray}   
where we used again  $\rho= 3n_s$ and $\delta=0.4$ MeV/$c^2$, which at intermediate densities are acceptable values to estimate the order of the number of self-gravitating APs at different $T$. It is easy to see from (\ref{K-N Formula}) that for $T=10^5$K, $T=10^7$K, and $T=10^8$K, one has $N^{sg}_{AP}\geqslant 10^{55}$, $N^{sg}_{AP} \geqslant 10^{58}$, and  $N^{sg}_{AP}\geqslant 10^{59}$ respectively, which means that in this range of temperatures the number of  APs needed to be self-gravitating is always much larger than the Chandrasekhar number, $N^{sg}_{AP} > N_{AP}^{Ch}$. Considering that for a single GRB event an estimate of about at least $10^{55}-10^{58}$ APs can be generated inside the bare strange star, and taking into account that the number of APs is conserved \cite{EJF-ViC-NPB}, one can infer that as the star cools down to temperatures equal to or below the threshold $T=10^7$K, the number of APs already inside the star will be enough to be self-gravitating and larger than the Chandrasekhar number $N_{AP}^{Ch}$ at some temperature between $10^5-10^7$K at which the star will collapse. Furthermore, given the enormous $\gamma$-ray activity in the galaxy center, there likely be more than one event hitting the young star, thus increasing the chances for the collapse to be induced at even higher temperatures. This AP-induced collapse scenario could then serve as an alternative mechanism to explain the lack of old pulsars in the galactic center. Our argument is consistent with the expectation that the missing pulsar problem is due to short-lived magnetars being created and blown away in the GC, rather than long-lived ordinary pulsars \cite{Dexter}.

Our scenario has some similarities but also significant differences with an alternate explanation for the missing pulsar problem that has been suggested in the literature \cite{DM, Ch-Bosons-2, Dermott}. While both scenarios are based on matter accumulation in the star center that reaches a limit after which a collapse is induced, they drastically differ in the origin of the matter that produces the collapse. The mechanisms in the literature are based on the dark matter (DM) halos that fill out the GC, or more precisely, on the capture of enough ambient DM by the pulsar to produce its collapse. The idea is that once the DM captured by the star is thermalized at the NS temperature, it starts to sink into the star center due to the gravitational pull, just as the AP is gravitationally attracted to the center in our AP mechanism. Nevertheless, in contrast to the DM scenario, the matter that accumulates in the star center in the case of the AP mechanism originates from the topological nature of matter-light interactions in the MDCDW quark matter phase. 

We underline that, as discussed in \cite{Fuller}, for the DM-collapse scenario to work, the DM density profile should satisfy a very restrictive constraint. Since the DM capture rate is proportional to the ambient DM density, the profile should be such that NS in the solar system neighborhood of $r\sim 10^4$ pc should survive for more than $10^{10}$ yrs, while those in the GC within a radius of $r=1$ pc need to collapse in at most $10^6$ yrs. This means that for the DM scenario to be viable, the DM density within the inner parsec needs to increase by at least a factor of $10^4$.

Therefore, the DM-induced collapse scenario constraints DM to a very stringent parameter space \cite{DM-Problems}. On the other hand, the AP scenario relies on: i) the apparent paradox of arguably high efficiency in magnetars formation from massive stars in the GC and their observed number scarcity \cite{Dexter}; ii) the existing evidence that several magnetars are associated with massive stellar progenitors ($M>40 M_\odot$) \cite{Figer}, which support the idea that magnetars formed in the GC are very massive compact objects made of quark matter; and iii) the well established observational fact \cite{GR} of a high GRB activity in the GC. All these elements combined support the viability of the AP mechanism proposed in this paper.

\section{Concluding remarks}

This paper discuss the viability of a new mechanism to explain the puzzling absence of pulsars in the GC. The main idea is that young magnetars formed in the region as quark stars in the MDCDW phase  will collapse under the bombardment of $\gamma$-rays. The mechanism is based on the fact that linear electromagnetic waves reaching the MDCDW medium can activate the chiral anomaly that couples the photons with the axions in the presence of a background magnetic field, and as a consequence, the waves can only propagate inside the quark medium as hybridized modes of APs \cite{EJF-ViC-NPB}. For application to magnetars, it is important to first check that the attenuation of the $\gamma$-rays does not prevent them from reaching the quark medium. In section 3, we showed that for hybrid starts this is not feasible, but for quark stars it is. Another challenge considered was whether, given the typical energy of the GRB and the properties of the MDCDW matter inside the star, the number of APs generated could be large enough for self-gravitation and eventually, for the generation of a mini black hole in the center. The analysis of section 4 confirmed this is indeed the case, so the star's collapse is unavoidable under these conditions.  

Since the AP mechanism is based on the existence of the MDCDW phase in the core of NSs, a valid question is how feasible this phase is as a matter structure candidate for NSs. A series of recent studies have explored this same question using several well-known astrophysical constraints. So far, all of them have found the MDCDW phase compatible with the observations. For example, thanks to its topology, the MDCDW phase has been shown to be very robust and stable against phonon fluctuations at temperatures, fields, and densities consistent with NSs observations and simulations \cite{MDCDW-3}. Moreover, the heat capacity of a NS in the MDCDW phase is well above $C_V\gtrsim 10^{36}(T/10^8)$ erg/K \cite{PRD20} which is the lower limit expected from long-term observations of NSs' temperatures in the range from months to years after accretion outburst together with continued observations on timescales of years \cite {Cv-NS}.

 Our analysis considered that the pulsars formed in the GC are mostly magnetars. This is based on the observation of magnetar SGR J1745-29 \cite{Magnetar-Detect}, a fact that revealed that the failures to detect ordinary pulsars at low frequencies could not be explained by a strong interstellar scattering. In other words, this observation signals some other intrinsic deficit mechanism should be at play here.  We argue that such a mechanism may be the AP one discussed in this paper. 

 Pulsars with strong inner magnetic fields are crucial for the AP mechanism to work. First, because a strong magnetic field ensures the stability of the MDCDW phase against thermal fluctuations up to temperatures much higher than those of NS \cite{Ferrer-Incera19}; and second, because the AP gap is proportional to the field, implying that the larger the field, the larger the gap, so the Chandrasekar limit is reachable at smaller AP numbers within the same range of densities. According to recent studies that used magnetohydrodynamic simulations in full general relativity on rapidly rotating stars \cite{relativisticmagnetohydro}, and investigations of the star's stability for slowly rotating NSs \cite{Cardall}, magnetar inner fields can reach values $\sim10^{17}-10^{18}$ G.
     
All these features, combined with the intense $\gamma$-ray activity known to exist in the GC \cite{GR}, warrant the conditions needed for the collapse of short-lived magnetars and thus the solution to the MPP puzzle via the AP mechanism.

Acknowledgments: This work was supported in part by NSF grant PHY-2013222.

\end{document}